# Quantum metamaterials: entanglement of spin and orbital angular momentum of a single photon

Tomer Stav[1], Arkady Faerman[2], Elhanan Maguid[2], Dikla Oren[1], Vladimir Kleiner[2], Erez Hasman[2*] and Mordechai Segev[1**]

1. Physics Department and Solid State Institute, Technion, Haifa 3200003, Israel
2. Micro and Nanooptics Laboratory, Faculty of Mechanical Engineering, and Russell Berrie Nanotechnology Institute, Technion, Haifa 3200003, Israel.

*mehasman@technion.ac.il
**msegev@technion.ac.il

**Metamaterials have been a major research area for more than two decades now, involving artificial structures with predesigned electromagnetic properties constructed from deep subwavelength building blocks. They have been used to demonstrate a wealth of fascinating phenomena ranging from negative refractive index and epsilon-near-zero to cloaking, emulations of general relativity effects, and super-resolution imaging, to name a few. In the past few years, metamaterials have been suggested as a new platform for quantum optics, and several pioneering experiments have already been carried out with single photons. Here, we employ a dielectric metasurface to generate entanglement between spin and orbital angular momentum of single photons. We demonstrate experimentally the generation of the four Bell states by utilizing the geometric phase arising from the photonic spin-orbit interaction. These are the first experiments with entangled states with metasurfaces, and as such they are paving the way to the new area of quantum metamaterials.**

The field of nanophotonics gave rise to metamaterials by the incorporation of materials science, physics, nanotechnology and electrical engineering, to produce structures with new electromagnetic (EM) properties which cannot be found in nature. Metamaterials are engineered structured, assembled from multiple element of a scale smaller than the wavelength of external stimuli, providing a medium with unique EM response and functionalities such as negative refraction[1,2], invisibility cloaking[3], and even near-zero permittivity and permeability[4,5]. On-chip integration of these nano-fabricated three-dimensional structures requires reducing the dimensionality of metamaterials yielding metasurfaces, providing new opportunities. Metasurfaces consist of a dense arrangement of dielectric or metallic resonant subwavelength optical antennas[6–14]. The light-matter interaction of an individual nanoantenna affords substantial control over the local phase pickup[15], thereby enabling local control over refraction and reflection. Accordingly, the light scattering properties of the metasurface can be manipulated by tailoring the nanoantennas' material, size, and shape via antenna resonance shaping[9,13,14], or through the geometric phase concept[6,7,12]. These alleged wavefront manipulations have been obtained for classical light. Recently, a metallic metasurface was introduced to the field of quantum light[16] for the purpose of coherent perfect absorption of single photons. However, despite visionary ideas proposing the use of metamaterials as a platform for quantum information, they have never been demonstrated to generate, manipulate or control entangled states, which are at the heart of the field of quantum information.

Quantum information is a multidisciplinary field that attempts to harness the laws of quantum mechanics to manipulate, transmit and detect information. By exploiting fundamental concepts in quantum physics, such as superposition and quantum entanglement, quantum information offers ways to solve problems in reduced time-complexity. For example, Grover's search algorithm provides quadratic speedup in comparison to its classical counterpart[17], or Shor's algorithm[18] which runs in polynomial time, while even the best known classical algorithms for integer factorization run in sub-exponential time[19]. One of the many possible realizations of these quantum algorithms may be achieved is using photons. The relatively easy manipulation of single photons, even in situations where their particle nature emerges, makes the construction of optical quantum processing units very appealing.

Photons can be controlled with the same optical devices used for classical light, they maintain their orbital angular momenta properties and quantum correlations very well unless absorbed[20,21]. That is, they do not suffer from severe decoherence problems as the alternative platforms to quantum information do. Cumbersome macro-scale devices and setups have been utilized to create photon entanglement[22], also spin to orbital angular momentum (OAM) entanglement was achieved via q-plates based on liquid crystals[23]. However, for more complex realizations of quantum computing units, needed to perform real-world calculations, the ability to manipulate quantum states at the nano-scale level is necessary. Recent advancements in on-chip quantum photonic circuits have shown the benefits of having integrated entangled photon sources[24,25]. In using metasurfaces for experiments with quantum light, it is extremely important to minimize the loss. For that reason, metallic metasurfaces which rely on plasmonic surface modes that are electronic resonances in their nature, inherently exhibit high loss. Thus, instead of metals, we employ metasurfaces made of high-refractive index dielectrics, which do not involve plasmonic decoherence and loss. Moreover, the compatibility with CMOS technology in the fabrication process therefore seems essential in the path to having large-scale quantum computation devices. We rely on the recent realizations of Si-based metasurfaces with efficiencies close to 100%[14,26], which makes them excellent candidates for quantum optics and quantum information applications.

**Here, we perform the first experiment using a metasurface to generate quantum entanglement.** We demonstrate, in experiments, that a dielectric metasurface can generate entanglement between the spin and the orbital angular momentum (OAM) of photons (Fig. 1). This is achieved by utilizing the Pancharatnam-Berry phase (geometric phase), which provides a photonic spin-orbit interaction mechanism[27–29]. We show the generation of the four Bell states at >90% fidelity, which exemplifies the superb ability of metamaterials platforms to produce and control quantum states of light, thus paving the way to new avenues in integrated photonic quantum information.

Our experiment relies on transmitting single photons through a dielectric metasurface that creates the entanglement between the spin and the OAM of a single photon. For this purpose, we fabricate the Si-based geometric phase metasurface (GPM) depicted in Fig. 2a. In general, GPMs are designed for spin-controlled wavefunction shaping, and are composed of anisotropic nanoantennas, designed to perform as nano half-wave plates, that generate a local geometric phase delay. The space-variant spin-dependent phase profile corresponds to an orientation function $\varphi_g = -2\sigma_\pm \theta(x,y)$, which defines the geometric phase of the light passing through the metasurface at position $(x,y)$ for the different spin states of single photons $\sigma_\pm = \pm 1$ (right handed and left handed circular polarizations). Accordingly, $\theta(x,y)$ is the in-plane orientation angle of the nanoantenna. In order to design a GPM that entangles the photon's spin to its OAM, the nanoantenna orientations are chosen to be $\theta(r,\varphi) = \ell\varphi/2$, where $\varphi$ is the azimuthal angle and $\ell$ is the winding number, in our case $\ell = 1$. Therefore, the GPM adds or subtracts $\Delta\ell = \pm 1$, one quanta of OAM, depending on the sign of the spin, and performs spin-flip $|\sigma_+\rangle \leftrightarrow |\sigma_-\rangle$. Such a metasurface performs the unitary transformation:

$$|\sigma_\pm\rangle|\ell\rangle \xrightleftharpoons{GPM} |\sigma_\mp\rangle|\ell \pm \Delta\ell\rangle . \qquad (1)$$

Consider now a single photon with zero OAM, polarized linearly in the horizontal polarization incident upon the metasurface. The state of the incident photon is described by a superposition of spins (circular polarizations), as

$$|H\rangle|\ell = 0\rangle = \tfrac{1}{\sqrt{2}}\left(|\sigma_+\rangle + |\sigma_-\rangle\right)|\ell = 0\rangle . \qquad (2)$$

After passing through the metasurface, the state of the photon becomes (following Eq. 1):

$$\frac{1}{\sqrt{2}}\left(|\sigma_-\rangle|\ell=\Delta\ell\rangle+|\sigma_+\rangle|\ell=-\Delta\ell\rangle\right). \quad (3)$$

In a similar fashion, an incident photon with zero OAM and in the vertical polarization, described by $|V\rangle|\ell=0\rangle = \frac{1}{\sqrt{2}i}\left(|\sigma_+\rangle-|\sigma_-\rangle\right)|\ell=0\rangle$ is transformed by the metasurface into

$$\frac{1}{\sqrt{2}i}\left(|\sigma_-\rangle|\ell=\Delta\ell\rangle-|\sigma_+\rangle|\ell=-\Delta\ell\rangle\right). \quad (4)$$

The states described by Eqs. 3 and 4 are maximally entangled states encoded on a single photon. The entanglement here is between the spin and the orbital angular momentum degrees of freedom. Interestingly, both the spin and the OAM represent angular momentum. Nevertheless, for photons whose spatial wavefunction is paraxial, as in our case, the spin and the OAM are totally independent, and have Hilbert spaces of different dimensions[29]. Furthermore, from expectation value perspective, the total angular momentum is conserved; the incident state is of zero total angular momentum as well as the state emerging from the GPM.

Our experimental setting is shown in Fig 2b. We generate a single photon in the state $|H\rangle|\ell = 0\rangle$, by spontaneous parametric down conversion (SPDC) of a λ=407.7 nm laser beam passing through a Type-II collinear phase-matched BBO crystal, and pass the light through the GPM. The interaction with the metasurface results in a single photon in an entangled state. Note that the experimental conversion efficiency of the metasurface was measured to be 72%.

To show entanglement, full quantum state tomography (QST) is performed on the quantum state, and the density matrix is recovered. For this end, we utilize a spatial light modulator (SLM) to project the state onto different OAM basis elements, and a set of quarter wave plate (QWP), half-wave plate (HWP) and a linear polarizer (Pol.) to project the state onto different elements of the polarization basis. The list of measurements is described in Table 1. We use coincidence counts between the two detectors, such that the single photon source is heralded. For integration time of 10 seconds ~1000 coincidence counts were measured without any projections. From a total of sixteen different measurements for each of the Bell states, which form a tomographically complete set of measurements, we recover the density matrix using a maximum likelihood estimation algorithm[30]. Using this technique, we experimentally recover the density matrices of the first two bell states $|\Psi^\pm\rangle = \frac{1}{\sqrt{2}}(|\sigma_+\rangle|\ell=-1\rangle \pm |\sigma_-\rangle|\ell=1\rangle)$ with fidelity of 0.9250 and 0.9496 for $|\Psi^+\rangle$ and $|\Psi^-\rangle$, respectively (Fig. 3b). As commonly done in such experiments, we define the fidelity between the recovered $(\tilde{\rho})$ and theoretical $(\rho)$ density matrices by $F(\rho,\tilde{\rho}) = \text{Tr}\left(\sqrt{\tilde{\rho}^{1/2}\rho\tilde{\rho}^{1/2}}\right)$. By flipping the GPM (the winding number flips sign, and now $\Delta\ell = -1$) it performs the unitary transformation $|\sigma_\pm\rangle|\ell\rangle \xrightarrow{GPM} |\sigma_\mp\rangle|\ell \mp \Delta\ell\rangle$ which enables the generation of the remaining two bell states $|\Phi^\pm\rangle = \frac{1}{\sqrt{2}}(|\sigma_+\rangle|\ell=1\rangle \pm |\sigma_-\rangle|\ell=-1\rangle)$. We perform QST on these states with fidelity of 0.9274 and 0.9591 for $|\Phi^+\rangle$ and $|\Phi^-\rangle$, respectively (Fig. 3b). These results are in very good agreement with theory (Fig. 3a).

Our demonstration of entanglement via quantum metamaterials and the generation of the four Bell states provide the route for nano-photonic quantum information applications. We anticipate that metasurfaces will become a standard tool in future quantum optics, and will be used extensively in photonic quantum information systems. These ideas can be extended to implement hyper-entangled state generation by using multifunctional or multispectral metasurfaces[26]. This conceptual progress of generating and controlling quantum states via metamaterials lead to many new ideas and directions, ranging from using metasurfaces to entangle two photons of difference frequencies and OAMs to manipulating quantum states of photons emitted from quantum dots in an integrated fashion. We intend to these ideas through future experimental demonstrations.

**Figure 1**

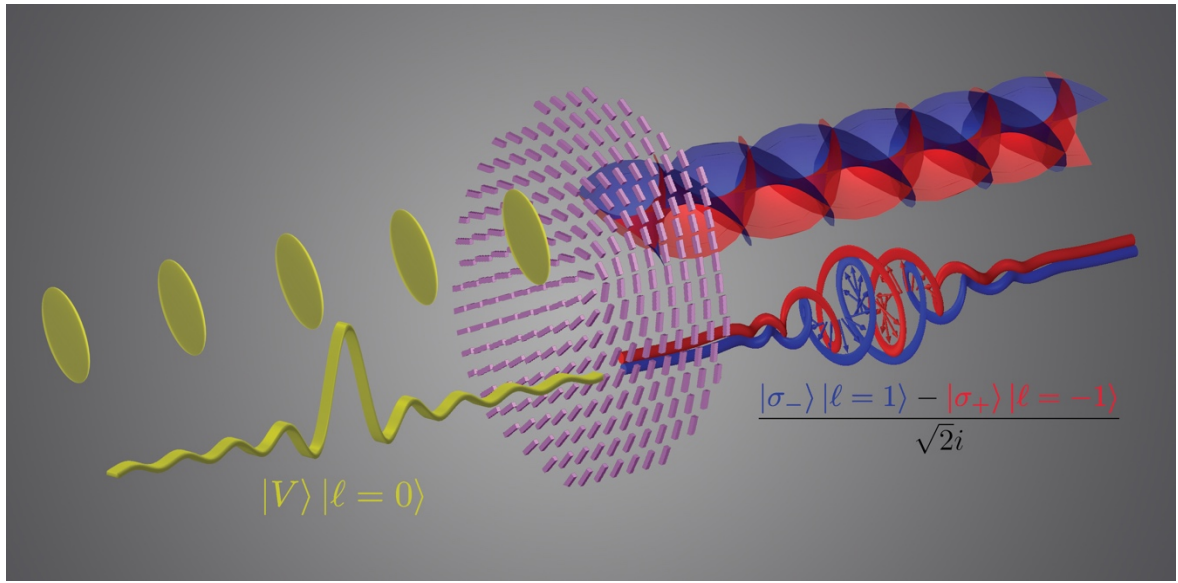

**Fig. 1. Illustrative depiction of the single photon entanglement.**

A single photon in vertical linear polarization is arriving from the left, as illustrated by the yellow electric field amplitude. This photon carries zero orbital angular momentum, as illustrated by the yellow flat phase fronts. The single photon passes through the metasurface comprising dielectric nano-antennae (purple), and exists as a quantum entangled state, depicted as a superposition of the red and blue electric field amplitudes and with the corresponding vortex phase fronts opposite to one another.

# Figure 2

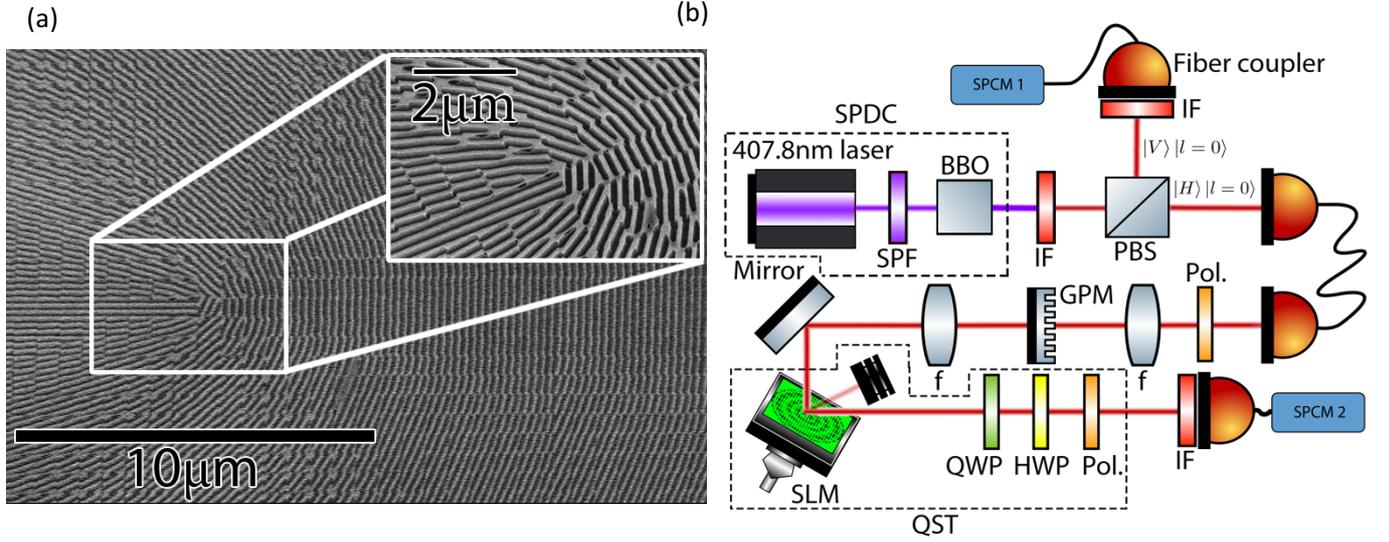

**Fig. 2. Experimental setup used to generate and measure the entangled states.**

(a) Scanning Electron Microscope (SEM) image of the Si-based GPM. Each building block in the GPM is composed of several nanorods, where their number is determined according to the orientation angle $\theta$ so as to fill an area of $700 \times 700$ nm$^2$. The nanorods are of 105 nm width and 300 nm depth, arranged 233 nm apart from each other. The metasurface diameter is of 200 μm. (b) Schematic of the experimental setup. A 407.8nm diode laser pumps a β-barium borate (BBO) crystal phase-matched for type-II collinear SPDC. The SPDC process produces two photons, one in vertical (V) polarization and the other in horizontal (H) polarization, centered around the degenerate wavelength of λ=815.6nm. The pump field and the photons produced at other wavelengths are filtered out by the interference filter (IF). The pairs of photons produced by SPDC are spatially separated using a polarizing beam splitter (PBS) and are coupled into fibers. The photon in V polarization acts as a trigger for the detection of the "signal photon" in H polarization (Eq. 2). The signal photon is coupled out of the fiber, passed through a linear polarizer (Pol.) and then passed through the GPM. The photon that has passed through the GPM is in a state that entangles spin and OAM. In the measurement process, this single photon is reflected off a phase-only SLM that projects the state onto different OAM basis. The reflected photons are projected on different polarization states by using polarization elements, and then measured in the Single Photon Counting Module (SPCM). Coincidence counts between the two SPCMs are used to measure different intensities for the QST.

| Order | Projection | QWP [deg] | HWP [deg] | SLM profile |
|---|---|---|---|---|
| 0 | Intensity | - | - | - |
| 1 | $|H\rangle|\ell=1\rangle$ | 0 | 0 | |
| 2 | $|V\rangle|\ell=1\rangle$ | 0 | 45 | |
| 3 | $|\sigma_+\rangle|\ell=1\rangle$ | 0 | 22.5 | |
| 4 | $|D\rangle|\ell=1\rangle$ | 45 | 22.5 | |
| 5 | $|D\rangle|\ell=-1\rangle$ | 45 | 22.5 | |
| 6 | $|\sigma_+\rangle|\ell=-1\rangle$ | 0 | 22.5 | |
| 7 | $|V\rangle|\ell=-1\rangle$ | 0 | 45 | |
| 8 | $|H\rangle|\ell=-1\rangle$ | 0 | 0 | |
| 9 | $|H\rangle|+\rangle$ | 0 | 0 | |
| 10 | $|V\rangle|+\rangle$ | 0 | 45 | |
| 11 | $|\sigma_+\rangle|+\rangle$ | 0 | 22.5 | |
| 12 | $|D\rangle|+\rangle$ | 45 | 22.5 | |
| 13 | $|D\rangle|r\rangle$ | 45 | 22.5 | |
| 14 | $|\sigma_+\rangle|r\rangle$ | 0 | 22.5 | |
| 15 | $|V\rangle|r\rangle$ | 0 | 45 | |
| 16 | $|H\rangle|r\rangle$ | 0 | 0 | |

**Table 1: list of measurements of the QST**.

The SLM phase profile range are from 0 (black) to $2\pi$ (white).

**Figure 3**

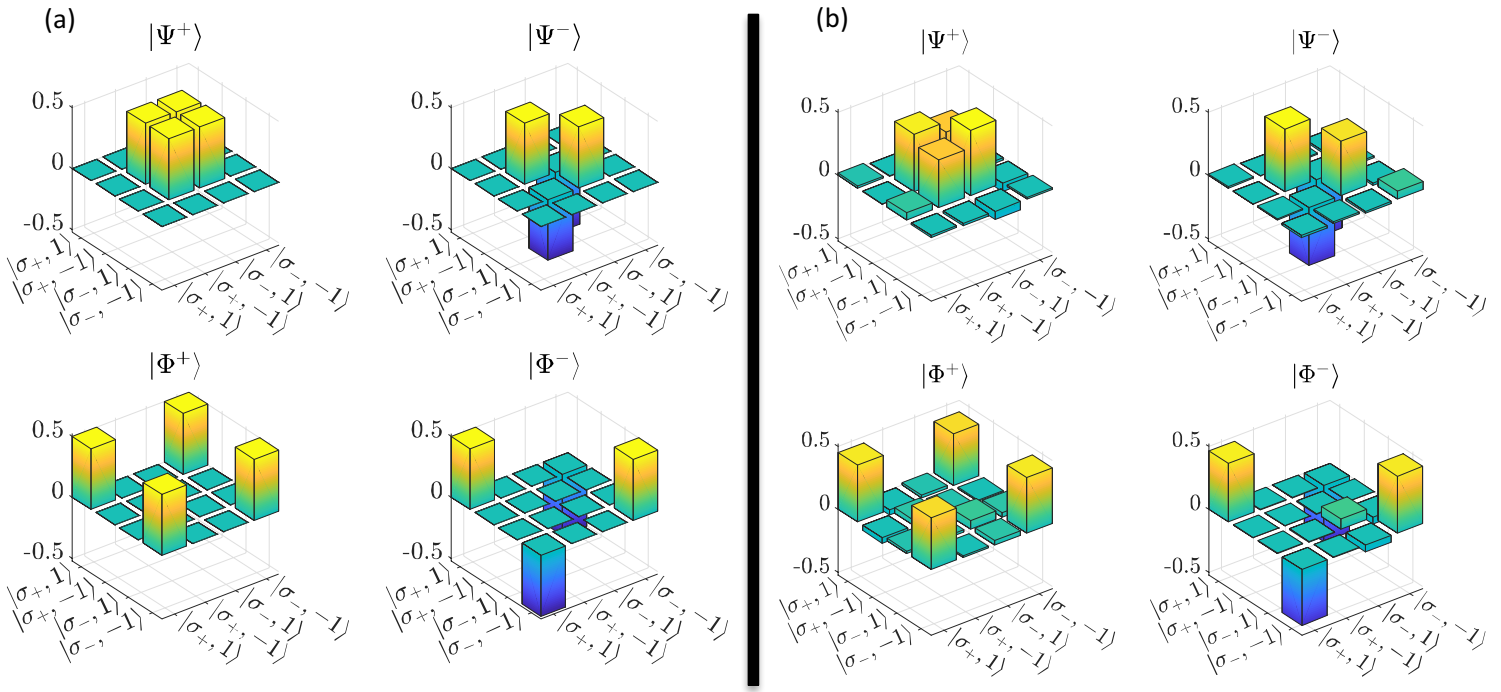

**Fig. 3. Density matrices of the four Bell states.**

(a) Theoretical density matrices for each of the Bell states. (b) Experimental density matrices measured for each of the Bell states using full quantum state tomography. The experimental results coincide with the theoretical results at fidelity higher than 90%.

Note: the results shown here are the real parts only, since the imaginary part is identically zero both theoretically and experimentally.